\documentclass[twocolumn, secnumarabic, amssymb, amsmath,nobibnotes, aps, prl,nofootinbib]{revtex4}
\begin{document}
\title{Geometric Entropies of Mixing (EOM)}
\author{B. H. Lavenda}
\email{bernard.lavenda@unicam.it}
\affiliation{Universit$\grave{a}$ degli Studi, Camerino 62032 (MC) Italy}
\date{\today}
\newcommand{\sumn}{\sum_{i=1}^{n}\,}
\newcommand{\sumk}{\sum_{i=1}^{k}\,}
\newcommand{\half}{\mbox{\small{$\frac{1}{2}$}}}
\newcommand{\fourth}{\mbox{\small{$\frac{1}{4}$}}}\newcommand{\twothirds}{\mbox{\small{$\frac{2}{3}$}}}
\begin{abstract}
 Trigonometric and trigonometric-algebraic entropies are introduced. Regularity increases the entropy and the maximal entropy is shown to result when a regular $n$-gon is inscribed in a circle. A regular $n$-gon circumscribing a circle gives the largest entropy reduction, or the smallest change in entropy from the state of maximum entropy which occurs in the asymptotic infinite $n$ limit. EOM are shown to correspond to minimum perimeter and maximum area in the theory of convex bodies, and can be used in the prediction of new inequalities for convex sets. These expressions are shown to be related to the phase functions obtained from the WKB approximation for Bessel and Hermite functions.
\end{abstract}
 
\maketitle
\section{Introduction}
In a companion paper \cite{L05}, EOM were related to polynomial and logarithmic functions. Here, we shall relate them to trigonometric and trigonometric-algebraic entropies. To the best of our knowledge trigonometric entropies were first discussed in reference \cite{B}, although we predicted their relation to EOM on the basis that regularity increases entropy. \par
In \cite{L05}, the EOM were derived from the distribution function using the Lorenz order. They were defined as the normalized difference between the dual of the  Lorenz function and the Lorenz function. The area between the two functions has been shown to be the Gini index of diversity \cite{K}.\par The limited applicability of Lorenz ordering in being able to derive  Lorenz curves from  probability distributions is well-known \cite{A}. In this paper we come across  examples where the Lorenz formulation fails. But, since it will be used to derive the relation between minimum perimeter in the theory of convex sets, we use the latter to derive the EOM associate with maximum area, thereby putting EOM on a geometrical basis. Moreover, the connections between EOM, extremal problems of convex sets, and the phase of the Bessel and Hermite functions in the WKB limit will brought out. This cross connection can be used to advantage in deriving new EOM from extremal problems of convex sets as well as predicting new inequalities between quantities characterizing convex sets. It will moreover be shown to furnish a geometric interpretation to the phases of orthogonal polynomials in the geometric optic limit.
\section{Isoperimetric theorems and EOM}
\subsection{Polygons inscribed}
We want to prove that the greatest EOM is the one in which of all the $n$-gons inscribed in a circle, the regular $n$-gon is the one with largest perimeter and area. The method of Lorenz function is applicable to this case.\par
Consider a set of independent random variables, $X_1,X_2,\ldots,X_n$, each having zero mean and unit variance such that the central limit theorem applies. Let $S_k=\sum_{i=1}^k X_i$, and let $N_n$ stand for the number of $S_k$'s that are positive, $1\le k\le n$, then in the limit
\begin{equation}
\lim_{n\rightarrow\infty}\Pr\left\{N_n/n<x\right\}=F(x)=\frac{2}{\pi}\sin^{-1}\sqrt{x}. \label{eq:arc}
\end{equation}
\par
The arcsine law (\ref{eq:arc}) is a symmetric distribution where the probabilities at the extremes, $x=0$, and $x=1$, are the greatest. Feller \cite{F} shows that the limiting distribution of the proportion of time a random walker spends on the positive half-axis has a cumulative distribution given by (\ref{eq:arc}). The central term has the smallest probability even though it coincides with the mean value, $m=\half$. This goes against intuition which equates mean and most probable values.\par
The Lorenz curve,
\[L(p)=p-\frac{\sin\pi p}{\pi},\]
and its dual,
\[\bar{L}(p)=p+\frac{\sin(1-p)\pi}{\pi},\]
reflect the symmetry of the arcsine distribution. Their difference gives the EOM 
\begin{equation}
S(p)=\half\left[\sin\pi p+\sin(1-p)\pi\right]=\sin\pi p, \label{eq:S-arc}
\end{equation}
upon normalization, which is concave and maximal. It is precisely the property of maximality, or that the EOM, as a measure of uncertainty out the outcome of an experiment, is greatest when all the outcomes have equal probabilities, that will be used to establish isoperimetric inequality for the inscription of $n$-gon into a circle based on Jensen's inequality for concave functions.\par The EOM, (\ref{eq:S-arc}), is easily generalized to a set of $n$ probabilities, viz.,
\[
S(p)=\frac{1}{n}\sumn \sin\pi p_i, \]
where the set $\{p_i\}$ is assumed to form a complete distribution. Since $(\sin\vartheta)^{\prime\prime}<0$ and $0<\vartheta<\pi$, Jensen's inequality gives \cite{HLP}
\begin{equation}
\sin\left(\pi\sumn p_i\big/n\right)>\frac{1}{n}\sumn\sin\pi p_i, \label{eq:Jensen}
\end{equation}
unless all the $p_i$ are equal. Now since the $p_i$ are assumed to form a complete distribution, (\ref{eq:Jensen}) reduces to
\begin{equation}
n\sin\left(\frac{\pi}{n}\right)>\sumn\sin\pi p_i. \label{eq:Jensen-bis}
\end{equation}
\par
The left side of (\ref{eq:Jensen-bis}) is half the perimeter of a regular polygon of $n$ sides inscribed in a circle of unit radius. Let $O$ be the center of the circle and $P_0,P_1,\ldots,P_n$  the vertices of a polygon that lie on the circle, where $P_0=P_n$ fixed but $P_1,P_2,\ldots,P_{n-1}$ can vary.  If the angle $P_{i-i}OP_i$ is identified with $\pi p_i$, then (\ref{eq:Jensen-bis}) asserts that both the perimeter and area of the polygon are greatest when the polygon is regular, viz., $P_{i-1}P_i=n$ for $i=1,2\ldots,n-1$. Hence, the familiar maximal properties of regular polygons coincide with the greatest EOM which occurs when all the probabilities are equal, $P_{i-1}OP_i=\pi/n$ for $i=1,2,\ldots,n-1$.\par
\subsection{Polygons circumscribed}
The Lorenz method fails to relate an EOM with the smallest perimeter and area of a $n$-gon of all the $n$-gons circumscribing a circle. The problem appears to be due to the fact that the underlying probability distribution has no moments.\par
Consider a nonregular $n$-gon circumscribing a circle of unit radius. The radii from the points of tangency of the sides of the $n$-gon to the circle subtend central angles $2\vartheta_i$. If the half-lengths of the segments are $\sqrt{\ell_i}$, then they will be related to their corresponding half-angles by
\[\vartheta_i=\tan^{-1}\sqrt{\ell_i}.\] These angles can vary from $0$ to $\pi$, and because their distribution is assumed uniform, their probabilities are
\[p_i=\frac{\vartheta_i}{\pi}=\frac{1}{\pi}\tan^{-1}\sqrt{\ell_i}.\]
This is the Cauchy distribution which has no moments.
\par
The inverse distribution function,
\[F^{-1}(\ell_i)=\tan^2\pi p_i,\]
would lead to a Lorenz curve
\[L(p_i)=\frac{1}{\pi}\tan\pi p_i-p_i,\]
which, although is positive and non decreasing, does not meet the other requisites of a Lorenz curve; in particular,  its dual is essentially the same as the Lorenz curve.\par
We cannot even associate the negative of the above function with  entropy since it is negative. However, the circumscribed $n$-gon can be looked on as a constraint which vanishes in the limit as $n\rightarrow\infty$. On account of this constraint there will be a reduction in entropy by the amount \cite{L95}
\[
\Delta S(p)=\sumn\left(p_i-\frac{1}{\pi}\tan\pi p_i\right). \]
 Jensen's inequality
\begin{eqnarray}
\Delta S(p)& = & 1-\frac{1}{\pi}\sumn\tan\pi p_i\nonumber\\
& \le & 1-\frac{n}{\pi}
\tan\frac{\pi}{n}=\Delta S(1/n),\label{eq:J}
\end{eqnarray}
holds if the $p$'s satisfy $0\le p_i\le\half$, or no angle can be greater than 90$^0$. This implies that $n>2$, for, otherwise, a polygon could not be circumscribed about the circle. Moreover, in the asymptotic limit
\[\lim_{n\rightarrow\infty}\Delta S=0,\]
and the system returns to its unconstrained state, the circle.\par
From Jensen's inequality (\ref{eq:J}) we conclude that the entropy reduction will be  largest---corresponding to the smallest entropy change---for the regular $n$-gon circumscribing a circle, which has the smallest perimeter and area among all $n$-gons. 

\section{Extrema of Convex Bodies and EOM}
\subsection{Minimum perimeter}
A convex body is characterized by its area $A$, perimeter $C$, diameter $D$, and thickness $E$ \cite{BF}. Inequalities between pairs of these quantities are well-known \cite{S52}. Three of the most common are
\begin{equation}
D\ge E\;\;\;\;\;\;\;\; C\ge\pi E, \;\;\;\;\mbox{and} \;\;\;\;\;\;\;\; \pi D\ge C. \label{eq:orbiform}
\end{equation}These inequalities are sharp because there exists bodies for which the equality  holds. In this case they are known as orbiforms which have no edges and each major chord is a double normal. The simplest orbiform that is not a circle is a Reuleaux triangle which has the same width in all directions. \par
Furthermore, inequalities also exist that involve more than two of these quantities. The inequality that we will have need of is
\begin{equation}
2\left\{\sqrt{D^2-E^2}+E\sin^{-1}\frac{E}{D}\right\}\le C. \label{eq:K}
\end{equation}
Consider a circle of radius $E$. Tangents to the circle from an outside point are equal, and the left hand side of
\begin{equation}
2\left\{\sqrt{D^2-E^2}-E\vartheta\right\}\le C-\pi E, \label{eq:K-bis}
\end{equation} 
where 
\begin{equation}
\vartheta=\cos^{-1}\frac{E}{D}, \label{eq:theta}
\end{equation} represents twice the length of the tangents less the arc  on the circle between the points of tangency. The minimum length (\ref{eq:K}) has also been the object of quantization, where the width $E$ takes on discrete values $n+\half$, with $n=0,1,2,\ldots$, and $D$ is the continuous radial coordinate \cite{KR}.\par Kubota \cite[p. 87]{BF} has shown that the equality in (\ref{eq:K}) occurs by appending two caps onto a circle of diameter $2E$ whose vertices lie on a line through the center of the circle at a distance D from the center of the circle.
\par
On the strength of the first inequality in (\ref{eq:orbiform}), inequality (\ref{eq:K}) can be written as
\[
\sin\vartheta-\vartheta\cos\vartheta\le\frac{1}{2D}\left(C-2\pi E\right).
\]
Doubling the right side and using the third inequality in (\ref{eq:orbiform}), result in 
\begin{equation}
\sin\vartheta-\vartheta\cos\vartheta\le\pi(1-\cos\vartheta).\label{eq:ineq}
\end{equation}
\par
For $\vartheta$ acute, (\ref{eq:ineq}) can be written as
\[E^2(\tan\vartheta-\vartheta)\le \pi ED(1-\cos\vartheta)\le 2\pi A,\]
on the strength of the Kubota inequality $2A\ge DE$ for a convex body of area $A$. The left hand side is the area of a circular arc of radius $E$, and whose peak is a distance $D$ from the center of the circle \cite{S52}.\par
Returning to the fundamental inequality (\ref{eq:ineq}), and allowing $\vartheta$ to vary over the interval $[0,\pi]$ we define the probability distribution by
\begin{equation}
p=\frac{D-E}{2D}=\half\left(1-\cos\vartheta\right)=\sin^2\half\vartheta, \label{eq:p}
\end{equation}
The greatest and least breadth of a convex set are the diameter $D$ and width $E$, respectively \cite{San}. The probability should be proportional to their difference. However, the $2$ in the denominator of (\ref{eq:p}) allows for \lq negative\rq\ widths, which, in actual case, implies the existence of a diametrically symmetrical cap of positive width.\par
Alternatively, (\ref{eq:p}) can be rationalized as the probability that any given point on the sphere will be covered by a randomly placed cap is the ratio of the area of the spherical cap, $2\pi r^2(1-\cos\vartheta)$, where $\vartheta$ is the central angle,  to the total surface area $4\pi r^2$ of a sphere of radius $r$ \cite{Solomon}.\par
Hence,
\[F^{-1}(p)=2\sin^{-1}\sqrt{p}=\pi L^{\prime}(p),\]
where we put $m=\pi$. Integrating from $0$ to $p$ gives the Lorenz curve,
\begin{equation}
L(p)=\frac{2}{\pi}\left\{\sqrt{p(1-p)}-(1-2p)\sin^{-1}\sqrt{p}\right\}, \label{eq:L-p}
\end{equation}
while integrating from $1-p$ to $1$ gives its dual
\begin{eqnarray}
\bar{L}(p) & = & 1-L(1-p)\label{eq:Lbar-p}\\ & = & 1-\frac{2}{\pi}\left\{\sqrt{p(1-p)}+(1-2p)\sin^{-1}\sqrt{1-p}\right\}. \nonumber
\end{eqnarray}
\par
 The Lorenz curve (\ref{eq:L-p}) is a nondecreasing convex function with $L(0)=0$ and $L(1)=1$, whereas its dual, (\ref{eq:Lbar-p}), is a nondecreasing concave function with $\bar{L}(0)=0$ and $\bar{L}(1)=1$.\par
The two-dimensional EOM
\begin{eqnarray}
S(p) & = & \half\left(\bar{L}-L\right) \nonumber\\
& = & 1-\frac{2}{\pi}\left\{(1-2p)\left[\sin^{-1}\sqrt{1-p}-\sin^{-1}\sqrt{p}\right]\right.\nonumber\\
&  &\left. +  2\sqrt{p(1-p)}\right\} \nonumber\\
& = & p+\frac{2}{\pi}\left\{(1-2p)\sin^{-1}\sqrt{p}-\sqrt{p(1-p)}\right\}\label{eq:2DS}
\end{eqnarray}
is both algebraic and trigonometric, but nonparametric. The end point conditions $S(0)=S(1)=1$ are satisfied, and (\ref{eq:2DS}) is symmetric having a maximum at $S(\half)=1-2/\pi$, thereby justifying its name as an EOM.\par
In terms of quantities characterizing convex bodies, the EOM (\ref{eq:2DS}) is seen to be the normalized minimum perimeter 
\begin{eqnarray}
\lefteqn{2\pi D\cdot S(E)}\nonumber\\ & = &  \pi( D-E)-2\left(\sqrt{D^2-E^2}- E\cos^{-1}\frac{E}{D}\right)\nonumber\\ 
&= & \pi D-2\left(\sqrt{D^2-E^2}+E\sin^{-1}\frac{E}{D}\right)\nonumber\\
&\ge &\pi D-C. \label{eq:S-geo}
\end{eqnarray} 
The second line is the sum of half of the difference between the circumferences of concentric circles with radii $D$ and $E$, and the wavefronts $\tau_{\pm}=\mbox{const.}$, where
\begin{equation}
\tau_{\pm}=\pm\left(\sqrt{D^2-E^2}-E\cos^{-1}\frac{E}{D}\right).
\label{eq:eikonal} 
\end{equation}
\par
These wavefronts are involutes to the circle of radius $E$, and  are chosen such that they travel in a counter clockwise direction \cite{KR}. The circle of radius $E$ is a caustic for these rays.  The rays corresponding to the eikonal $\tau_{-}$ are half-lines tangent to the circle of radius $E$. From the half arc length $E\cos^{-1}(E/D)$, the length $\sqrt{D^2-E^2}$ must subtracted because the direction of the ray is from the outer to the inner circle. The outgoing wavefront $\tau_{+}$ has the signs reversed.\par The closed curve, or cap, therefore consists of a ray from the caustic to the boundary,  a reflected wave from the boundary to the caustic, and an arc on the caustic between the two points of tangency. Apart from the phase change as the wavefront passes through the caustic, (\ref{eq:eikonal}) is the asymptotic expression for phase of the ordinary Bessel function in the periodic region \cite{BH}.\par
The   penultimate line in (\ref{eq:S-geo}) is the difference between the circumference of a circle of diameter $D$ and the perimeter of the two symmetrically placed caps joined by an arc of a circle of radius $E$. The EOM predicts the existence of a diametrically symmetric cap on the caustic, whose length is the convex hull of a circle and two points located outside for which the equality sign in (\ref{eq:S-geo}) holds \cite{S52}. The inequality in the last line of (\ref{eq:S-geo}) ensures that the perimeter of all other convex bodies will be greater. The EOM corresponds to that of a convex body with minimum perimeter.\par
In terms of the angle variable the Lorenz curve, (\ref{eq:L-p}), and its dual, (\ref{eq:Lbar-p}) are
\begin{equation}
L(\vartheta)=\frac{1}{\pi}\left(\sin\vartheta-\vartheta\cos\vartheta\right), \label{eq:L-t}
\end{equation}
and
\begin{eqnarray}
\bar{L}(\vartheta) & = & 1-L(\pi-\vartheta)\label{eq:Lbar-t}\\  & = & 1-\cos\vartheta-\frac{1}{\pi}\left(\sin\vartheta-\vartheta\cos\vartheta\right), \nonumber
\end{eqnarray}
respectively. The nonnegativity of  (\ref{eq:L-t}) is guaranteed by a fundamental trigonometric inequality, while the nonegativity of  (\ref{eq:Lbar-t}) rests on our fundamental inequality (\ref{eq:ineq}) for a convex set.\par
Therefore, the EOM (\ref{eq:2DS}), as a function of angle, is 
\begin{equation}
S(\vartheta)=\half\left(1-\cos\vartheta\right)-\frac{1}{\pi}\left(\sin\vartheta-\vartheta\cos\vartheta\right), \label{eq:2DS-bis}
\end{equation}
with $S(0)=S(\pi)=0$. The EOM is maximal at $\vartheta=\half\pi$, which can be considered as the asymptotic limit as $D\rightarrow\infty$, with $E$ fixed.  Inequality (\ref{eq:K}) reduces to $2D\le C$, where the equality sign is satisfied by  a secant passing through the center \cite{S52}. For values of $\vartheta>\half\pi$, the thickness becomes \lq negative\rq. This is not unlike the existence of \lq negative\rq\ temperatures where the entropy as a function of the energy is bell-shaped curve and states on the descending part of the curve were associated with states of \lq negative\rq\ temperature \cite{L99}.\par
However, there are two solutions
\begin{equation}
\cos^{-1}\frac{E}{D}=\vartheta_{\pm} \label{eq:cos}
\end{equation}
corresponding to two simple saddle points in the method of steepest descents \cite{BH}, where $0<\vartheta_{+}<\half\pi$, and $\half\pi<\vartheta_{-}=\pi-\vartheta_{+}<\pi$. This is the origin of the symmetric cap whose vertices lie on a straight line through the center of the circle, and whose central angle is \begin{equation}\vartheta_{-}=\cos^{-1}(-E/D)=\pi-\cos^{-1}(E/D). \label{eq:cos-bis} 
\end{equation}
The state of the greatest EOM corresponds to the coalescence of the two saddle points, 
\[\lim_{D\rightarrow\infty}\cos^{-1}\frac{E}{D}=\half\pi.\]  
The asymptotic expression for the Lorenz curve,
\begin{equation}
L=\frac{2D-\pi E}{2\pi D}, \label{eq:L-asy}
\end{equation}
 requires $D>\pi E/2$, and is stronger than the first inequality in (\ref{eq:orbiform}).\par
The term within the parenthesis on the left side in (\ref{eq:K}) is precisely the WKB expression for the phase of the Bessel function in the periodic region, $D>E$ \cite{L}. This requires that the order $E$ be less than the argument, $D$, of the Bessel function. The phase in the transition region from periodic to exponential solutions is given by (\ref{eq:L-asy}) \cite{Debye}.
\par
\subsection{Maximum area}
The maximum area of a convex body satisfies the inequality 
\begin{equation}
E\sqrt{D^2-E^2}+D^2\sin^{-1}\frac{E}{D}\ge 2A. \label{eq:A}
\end{equation}
The equality sign pertains to a body formed by removing from a circle points outside two symmetrically placed secants \cite{S52}.\par
Rearranging the left  side  of (\ref{eq:A}) gives
\[
2A\le 
\half\pi D^2-\left(D^2\cos^{-1}\frac{E}{D} -  E\sqrt{D^2-E^2}\right). \]
This appears in the problem of overlapping circles. If two circles of equal radius, $\half D$, have their centers at a distance $E$ apart, which is less than $D$, they will overlap in a region which has been referred to as a \lq lens\rq \cite{BF} or \lq lune\rq \cite{KM}, and  whose area is
\begin{eqnarray}
A_{o} & = & \half D^2\int_0^{\cos^{-1}(E/D)}\sin^2\vartheta\,d\vartheta\nonumber\\\
& = & \fourth D^2\left\{\cos^{-1}\frac{E}{D}-\frac{E}{D}\sqrt{1-\frac{E^2}{D^2}}\right\}. \label{eq:A-over}
\end{eqnarray}
The resulting area in (\ref{eq:A}) has a crescent shape. The maximum area in (\ref{eq:A}) is attained for this \lq symmetric lens\rq, symmetric referring to equal radii. We will now see how the corresponding EOM can even lead to sharper results.\par
Subtracting $\half E^2$ from both sides of (\ref{eq:A}), and
 dividing by $\half\pi D^2$ give
\begin{eqnarray*}
S(D)   
& = &  1-\frac{E^2}{D^2}-
\frac{2}{\pi}\left(\cos^{-1}\frac{E}{D}
 -  \frac{E}{D}\sqrt{1-\frac{E^2}{D^2}}\right)
\\
& \ge & \frac{4A-\pi E^2}{\pi D^2}. 
\end{eqnarray*}
The second line associates the EOM with the maximum area of a convex body. In terms of angle and probability, the EOM is
\begin{eqnarray*} S & = & \sin^2\vartheta-\frac{1}{\pi}\left(2\vartheta-\sin2\vartheta
\right) \\
& = & p-\frac{2}{\pi}\left(\sin^{-1}\sqrt{p}-\sqrt{p(1-p)}\right),
\end{eqnarray*}
where  the probability that a point will fall in the area of a ring of width $D-E$ is
\begin{equation}
p=\frac{\pi(\half D\sin\vartheta)^2}{\fourth\pi D^2}=\frac{D^2-E^2}{D^2}. \label{eq:p-bis}
\end{equation}
On account of (\ref{eq:theta}), it is also the probability of $\vartheta$, whose density is $2\sin\vartheta\cos\vartheta$, $0\le\vartheta\le\half\pi$.\par
The EOM  is bell-shaped but non-symmetric, having its zeros at $D=E$ and $D\rightarrow\infty$. It has a maximum at $p=\pi/(\pi+4)\approx 0.7$, corresponding to $D\approx 2E$.  \par
The maximum area in (\ref{eq:A}) is precisely the phase of the Hermite polynomial in the WKB limit \cite[p. 392]{L}. In contrast to Bessel functions, where $E$ is quantized and $D$ continuous, the width $E$ corresponds to the oscillator displacement, and $D$ is the eigenvalue of Hermite's differential equation, $D=\sqrt{2n+1}$, $n=0,1,2,\ldots$. \par
\section{Conclusions}
Due to the limited applicability of Lorenz ordering in the derivation of EOM from probability distributions, a geometric approach has been used. New trigonometric and algebraic entropies have been introduced on this basis. In  cases where Lorenz ordering is valid, the greatest trigonometric EOM has be shown  to be associated with the regular $n$-gon inscribed in a circle, showing that regularity increases entropy. Lorenz ordering has also led to a trigonometric-algebraic EOM associated with the minimum length of a convex body. This has further been shown to be related to the phase of a Bessel function in the WKB limit.\par
Lorenz ordering cannot be used to establish the connection between largest entropy and the smallest perimeter and area of an $n$-gon circumscribing a circle. Rather, the problem of circumscription has been looked upon as a constraint with the constraint vanishing in the limit as $n\rightarrow\infty$. The entropy reduction has been calculated  for the circumscription, and is greatest when all the sides are equal. This implies minimum entropy change.\par
Finally, the maximum area of a convex body has been tackled where it is shown that the EOM is associated with the maximum area of a convex body. This corresponds to the phase of a Hermite polynomial. Hence, there is a cross fertilization among EOM, isoperimetric theorems, extremal problems in convex sets, and the phases of orthogonal polynomials in the geometric optic limit.


\begin{thebibliography}{99}
\bibitem{L05}B. H. Lavenda, \lq Lorenz ordering and entropies of mixing\rq, submitted to \emph{Open Systems \& Information Dynamics\/}.
\bibitem{B}M. Behara, \emph{Additive and Nonadditive Measures of Entropy\/} (Wiley, New Delhi, 1990) Part 2.
\bibitem{K}G. Koshevoy and K. Mosler, \emph{J. Am. Statist. Ass.\/} {\bf{91}} (1996) 873.
\bibitem{A}B. C. Arnold, \emph{Majorization and the Lorenz Order: A Brief Introduction\/} 
(Springer-Verlag, Berlin, 1987).
\bibitem{F}W. Feller, \emph{Introduction to Probability and Its Applications\/} vol. I, 3rd ed. (Wiley, New York, 1968) p. 81.
\bibitem{HLP}G. Hardy, J. E. Littlewood, and G. P\'olya, \emph{Inequalities\/} 2nd ed. (Cambridge
U. P., Cambridge, 1952) p. 99.
\bibitem{L95}B. H. Lavenda, \emph{Thermodynamics of Extremes\/} (Horwood, Chichester, 1995).
\bibitem{BF}T. Bonnesen and W. Fenchel, \emph{Theory of Convex Bodies\/}, transl. (BCS Associates, Moscow (Idaho), 1987) p. 87.

\bibitem{S52}M. Scholander, \emph{Trans. Amer. Math. Soc.\/} {\bf{23}} (1952) 139.
\bibitem{KR}J. B. Keller and S. I. Rubinow, \emph{Ann. Phys.\/} (NY) {\bf{9}} (1960) 24.
\bibitem{San}L. A. Santal\'o, \emph{Integral Geometry and Geometric Probability\/} (Addison-Wesley, Reading MA, 1976) p. 6.

\bibitem{Solomon}H. Solomon, \emph{Geometric Probability\/} (SIAM, Philadelphia, 1976) p. 89.
\bibitem{BH}N. Bleistein and R. A. Handelsman, \emph{Asymptotic Expansions of Integrals\/} (Dover, New York, 1986) p. 269.
\bibitem{L99}B. H. Lavenda, \emph{J. Phys. A: Math. Gen.\/} {\bf{32}} (1999) 4297.
\bibitem{L}C. Lanczos, \emph{Linear Differential Operators\/} (Van Nostrand, London, 1961) p. 395.
\bibitem{Debye}P. Debye, \emph{Mathematische Annalen\/} (1910) 535.

\bibitem{KM}M. G. Kendall and P. A. P. Moran, \emph{Geometrical Probability\/} (Griffin, London, 1963) p. 107.

\end{thebibliography}
\end{document}